\documentclass[lettersize,journal]{IEEEtran}
\IEEEoverridecommandlockouts
\usepackage{cite}
\usepackage{soul}
\usepackage{amsmath,amssymb,amsfonts}
\usepackage{booktabs,xcolor}
\usepackage{url}
\usepackage[english]{babel}
\usepackage{blindtext}
\usepackage{wrapfig}
\usepackage{xfrac}
\usepackage[tight]{subfigure}

\usepackage[ruled, lined, linesnumbered, commentsnumbered, noend ]{algorithm2e}
\usepackage[noend]{algpseudocode}
\usepackage{longtable}

\newcommand\systemname{\textsc{RIFO}\xspace}

\newcommand{\ToN}[1]{\textcolor{black}{#1}}
\newcommand{\ToNMinor}[1]{\textcolor{black}{#1}}
\def\BibTeX{{\rm B\kern-.05em{\sc i\kern-.025em b}\kern-.08em
    T\kern-.1667em\lower.7ex\hbox{E}\kern-.125emX}}
\usepackage{balance}

\begin{document}
\sloppy
\title{RIFO: Pushing the Efficiency of\\Programmable Packet Schedulers}
\author{Habib Mostafaei\thanks{Habib Mostafaei,  Eindhoven University of Technology,
The Netherlands}, Maciej Pacut\thanks{Maciej Pacut , TU Berlin, Germany}, Stefan Schmid\thanks{Stefan Schmid,  TU Berlin and Fraunhofer SIT, Germany}}

\maketitle

\begin{abstract}
Packet scheduling is a fundamental networking task that recently received renewed attention in the context of programmable data planes.
Programmable packet scheduling systems such as those based on Push-In First-Out (PIFO) abstraction enabled flexible scheduling policies, but are too resource-expensive for large-scale line rate operation.
This prompted research into practical programmable schedulers (e.g., SP-PIFO, AIFO) approximating PIFO behavior on regular hardware.
Yet, their scalability remains limited due to extensive number of memory operations. 
To address this, we design an effective yet resource-efficient packet scheduler, Range-In First-Out (RIFO), which uses only three mutable memory cells and one FIFO queue per PIFO queue. RIFO is based on multi-criteria decision-making principles and uses small guaranteed admission buffers. Our large-scale simulations in Netbench demonstrate that despite using fewer resources, RIFO generally achieves competitive flow completion times across all studied workloads, and is especially effective in workloads with a significant share of large flows, reducing flow completion time up to 4.91x in datamining workload compared to state-of-the-art solutions. Our prototype implementation using P4 on Tofino switches requires only 650 lines of code,  is scalable, and runs at line rate.
\end{abstract}


\section{Introduction}\label{sec:intro}

\IEEEPARstart{P}{acket} scheduling is a fundamental function of network switches. By optimizing  the timing and order of forwarded packets, 
a scheduler can greatly improve performance metrics such as flow completion times (FCT), throughput, and tail latency, as well as QoS guarantees and fairness~\cite{ tailLatency-sigcomm92, pFabric-sigcomm13, fairQueueing-sigcomm89}. 

Programmable packet schedulers~\cite{UniversalPacketScheduling, pifo-sigcomm16, TowardsProgrammable} recently received much attention because of their flexibility and for allowing network operators to implement their scheduling policies, even on a per-packet level, without changing the hardware design. Programmable scheduling consists of two parts: a policy that assigns priorities (\emph{ranks}) to packets, and a scheduler that favors packets with low ranks over the packets with high ranks.
By ranking, reordering, and dropping packets, an effective scheduling algorithm may realize various performance goals depending on its policy. For example, assigning ranks according to the remaining flow size results in reduced flow completion times~\cite{srpt-OR66}.

The performance of the system depends on the scheduler: the part of the system that processes streams of packets with ranks pre-assigned according to the policy of choice.
Originally, programmable schedulers were designed for the Push-In-First-Out~(PIFO) scheduler~\cite{pifo-sigcomm16}, where incoming packets are sorted by ranks: the scheduler maintains the queue in sorted order of packet ranks, and packets are dequeued from the head.
This natural scheduling closely realizes the goals set by the rank-assigning policy, but packets need to be inserted in a strict order, which entails high computational costs.

To support non-programmable schedulers, researchers investigated scalable alternatives to PIFO, which can be performed at line rate.
By approximating only some aspects of PIFO while ignoring the others, some schedulers demonstrated that one can still closely realize the goals of the rank-assigning policy at a much lower computational cost.
For example, Strict-Priority PIFO~(SP-PIFO)~\cite{sp-pifo-nsdi20} focused on forwarding the packets in the correct order using multiple strict priority queues, and Admission-In First-Out (AIFO)~\cite{aifo-sigcomm21} focused on admitting the right packets using a FIFO queue.





This paper complements this line of research, which primarily revolves around performance, by exploring computational efficiency, and especially considering the amount of memory and the required number of packet processing stages. 
In addition to reducing the resource footprint, this may also benefit performance gain: by reusing registers and reducing the number of operations and processing stages, packet processing latency can be improved~\cite{p8-TNSM21, tofino}.
The network operators can use the saved memory for other functions, such as buffering, forwarding, and filtering. They can also support a larger number of slices or tenants without deploying additional switches or upgrading existing ones. The utility of additional memory depends on the application, e.g.~\cite{ZhuWHPS022} reports on the effect of additional available memory on satisfaction levels (the fraction of a network application's lifetime that it meets its utility target) in heavy hitter detection, superspreader detection, maintaining TCP connection and caching, where doubling the available amount of memory improves satisfaction by 1.1-1.5x initially, however at a certain point adding memory results in diminishing returns.


We present Range-In First-Out (\systemname), a programmable packet scheduling algorithm that processes packets at line rate using only a single FIFO queue and three registers. \systemname focuses on admitting the right set of packets without reordering them. It classifies arriving packet ranks into small and large categories, similar to flow classification. To achieve this, we use \emph{min-max linear normalization}~\cite{Triantaphyllou00,Tzeng11} for its simplicity and efficiency.
To normalize, we maintain just two characteristics of the recent packets: the minimum and the maximum packet ranks, and periodically reset these values to avoid outliers in detecting the correct ranks.
To admit a packet, \systemname uses the score assigned by the normalization, and if the score is higher than the current queue utilization, it admits the packet. 
Furthermore, we use small guaranteed admission buffers; if our queue utilization is low, we admit packets irrespective of their scores.
We demonstrate that min-max normalization is an efficient alternative to approximate quantile~\cite{aifo-sigcomm21} in the data plane of programmable switches for packet admission decisions based on queue occupancy. Our normalization approach significantly reduces memory requirements, crucial for deployment in resource-constrained environments, as demonstrated by our lookup tables' compact footprint requiring $\approx$ 1.22 KB of memory.

\ToNMinor{ \systemname takes a more resource-efficient approach to admission control compared to state-of-the-art methods that use a single FIFO queue, such as AIFO. \systemname uses a resetting logic that periodically refreshes the minimum and maximum packet ranks, allowing recent traffic patterns to be captured with a simpler form of quantile estimation. This approach avoids AIFO's sliding windows and reduces memory usage by relying solely on min-max normalization with just three counters. As a result, RIFO achieves efficient packet admission and prioritization with minimal resources, providing a more lightweight alternative to AIFO for resource-constrained environments while maintaining comparable performance.}

RIFO's resource efficiency is pushed to an extreme by using only three mutable registers, and one FIFO queue. Despite that, our evaluations show that the performance of \systemname is robust, with high throughput for different workloads, and competitive with state-of-the-art scheduling solutions in terms of flow completion times and fairness. For workloads containing many large flows, such as datamining, our method can improve the FCT by up to 2.25x, 1.73x, and 4.91x compared to PIFO, SP-PIFO, and AIFO, respectively, depending on the traffic load of the links.


\subsection{Contributions}

We summarize our main contributions as follows:
\begin{itemize}
    \item We present \systemname, a programmable packet scheduler that is simple, resource-efficient, scalable, and implementable at line rate in existing programmable switches.
    It is based on multi-criteria decision-making normalization techniques, carefully chosen admission criteria, and uses small guaranteed admission buffers. \systemname maintains only two statistics about the recently seen packets' ranks ($Min$ and $Max$).
    
    \item Our large-scale simulations of \systemname report up to 4.91x lower FCT for large flows, when the workload contains a significant share of large flows (e.g., datamining workload), and robust performance across all workloads, compared with state-of-the-art programmable schedulers.
    Similar results hold for realizing fair queuing.
    \item We find that \systemname is robust across various rank distributions, by studying both our approach and existing programmable schedulers under more traffic traces compared to other admission-based packet schedulers. This sheds light on the performance of scheduling algorithms in various conditions.
    \item \ToN{We present a proof-of-concept implementation on Tofino switches~\cite{tofino}, using just five processing stages.
    Our comparisons with other implementations demonstrate improvements in hardware resource consumption metrics, 2.54x compared with AIFO and 6.55x with SP-PIFO.
     \item  To contribute to the research community and ensure reproducibility, we will make our source code and experimental artifacts publicly available alongside this paper.}
\end{itemize}

\subsection{Organization}

The remainder of this paper is organized as follows.  In \textsection\ref{sec:background}, we provide the basis of programmable packet schedulers. \ToN{ We introduce the design of \systemname in \textsection\ref{sec:algorithm}. In \textsection\ref{sec:analysis}, we analyze the accuracy of \systemname and \textsection\ref{sec:dataplane} reports its Tofino implementation and challenges.} 
\textsection\ref{sec:eval} reports on the performance of our empirical evaluation under different workloads, as well as on the resource consumption.
After putting our work into perspective with existing programmable schedulers in \textsection\ref{sec:relatedWork}, we conclude in \textsection\ref{sec:conclusion}.


\section{Background}\label{sec:background}

Programmable packet scheduling~\cite{pifo-sigcomm16} promises a flexible packet scheduling system able to realize a wide range of policies (e.g. with the goal of minimizing flow completion time or maximizing fairness) without the need for changing the hardware design when the policy changes.
The system decouples scheduling into (1) packet rank assignment from the packet fields and (2) rank-based scheduler that coincides packets with their ranks, ignoring the packet fields at that point.
The lower the rank, the more time critical is the packet.
The rank sums up all the scheduling-relevant information about the packet, so two packets with the same rank are treated as equally important, regardless of the differences in their fields.
This stands in contrast to the non-programmable packet scheduling, where no concept of a rank exists\footnote{Although in non-programmable packet scheduling, the packets have priorities, those are just one of the parameters in addition to the packet fields, rather than a single characteristic that summarizes the importance of the packet.}.

To effectively realize the goals of a rank-assigning policy of choice, the rank-based scheduler must prefer the packets with smaller ranks over the packets with higher ranks.
The pioneer programmable scheduling system Push-In-First-Out~(PIFO)~\cite{pifo-sigcomm16} proposes to \emph{sort} packets according to their ranks: PIFO uses a priority queue, inserting incoming packets in positions according to their ranks (with an exception that packets arriving when the queue is full are dropped).
When the link is idle, the system schedules the packet with the smallest rank.
We note that packets are not required to be stored in per-flow queues.

The flexibility of the system allowed the implementation of multiple scheduling policies.
The authors of PIFO report that their scheduler can realize many
existing non-programmable scheduling algorithms, e.g., Weighted Fair Queueing~\cite{DemersKS89}, Token Bucket Filtering~\cite{Token}, Hierarchical Packet
Fair Queueing~\cite{BennettZ97}, Least-Slack Time-First~\cite{Leung2005ANA}, the Rate-Controlled Service Disciplines~\cite{10.5555/2692227.2692232}, and fine-grained priority scheduling (e.g., Shortest Job First).
Some of these policies are straightforward to implement, e.g. setting ranks to the job size realizes Shortest Job First, setting ranks to the pre-assigned slack plus the arrival time realizes Least-Slack Time-First, whereas realizing more sophisticated policies requires maintaining additional state (see the PIFO paper for examples~\cite{pifo-sigcomm16}).
All these policies modify only the rank-assigning function, and a single rank-based scheduler can realize any policy.




\color{black}

\section{The Design of RIFO}\label{sec:algorithm}

Our goal is the design of simpler and more resource-efficient programmable scheduling algorithms 
which perform well under a variety of workloads.
Resource efficiency leaves the precious resources of the switch for other purposes, such as FIFO queues for multi-tenancy, and registers for load balancing, diagnostics, or routing.
In addition to reducing the resource footprint, resource efficiency may also decrease packet processing latency by reducing the number of registers, operations, and packet processing stages~\cite{p8-TNSM21, tofino}.

\subsection{Rationale behind \systemname}
\noindent\textbf{Rank classification}. 
The design of \systemname is based on the classification of the packet ranks using normalization. \systemname classifies packets based on their rank values computed by the programmable packet schedulers into a relative rank value that identifies the corresponding size of the flow it belongs to compared with the others:
\ToN{
essentially, a quantile computation for the incoming rank. Our rank classification does not involve computing the quantile exactly but rather approximates it using a linear function.}

This rank classification is similar to flow classification, which classifies the flows into small, medium, and large ones. \systemname leverages this classification to decide on the admission of the packets. 

\noindent\textbf{Rank admission}. 
After identifying the relative size of the arrived ranks, \systemname needs an admission mechanism to decide if it has to drop the packet. \systemname decides the admission of the packets by checking the queue occupancy of the egress port. We note that admitting more packets from flows with low ranks, i.e., the packets that belong to large flows, likely increases the dropping probability~\cite{bufferSizing-sigcomm04}. \systemname relies on a single decision criterion for the packet admission, and if there is room in the queue for the packets of large flows, it admits them. Otherwise, it prefers dropping the packets of large flows in each range.  Since the goal is to minimize the FCT of small flows, \systemname admits the packets with relatively small ranked values.

\subsection{The Algorithm}

RIFO uses a single FIFO queue and admits an incoming packet depending on the current queue utilization.
RIFO computes only two statistics about the recently seen packets' ranks: the minimum and maximum packet rank, which can be stored using only two registers.
Using these two statistics with a carefully chosen normalization function, we can position the rank of an incoming packet relative to the recently seen packets.
We identify that admitting high-ranked packets comes with a penalty for our system (possibly denying buffer space for low-ranked packets), hence we use a representative of the \emph{cost-criteria normalization functions} from multi-criteria decision-making optimization~\cite{Triantaphyllou00,Tzeng11,JAHAN2015335}. 
As we demonstrate in our evaluation, this minimalistic information is sufficient for efficiency and robustness.

\begin{figure}[ht]
    \centering
    \includegraphics[width=\columnwidth]{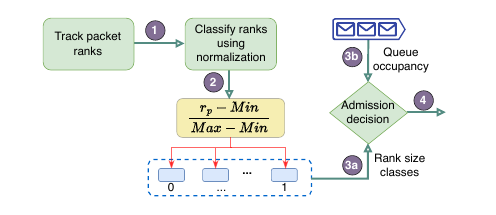}
    \caption{The general architecture of \systemname.}
    \label{fig:RIFO-architecture}
\end{figure}
 
\begin{figure*}[t]
    \centering
    \includegraphics[width=\linewidth]{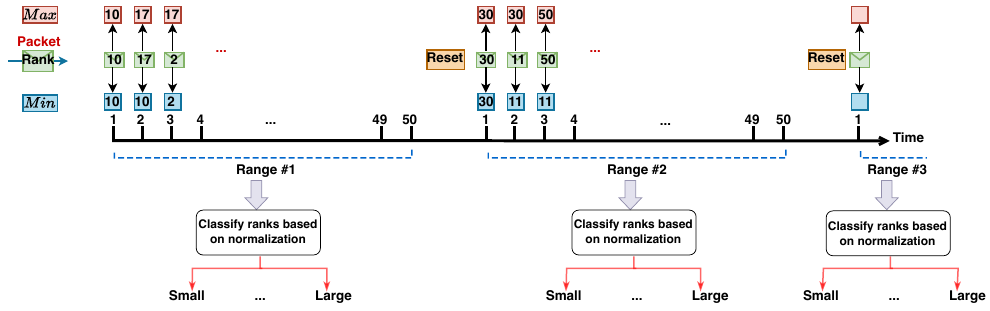}
    \caption{The track range resetting mechanism of \systemname with $T$=50. We maintain three registers: corresponding to the value of $Min$, $Max$ and the counter of packets seen since the last reset. Initially, we set the counter to $0$, and the values $Min$ and $Max$ to the rank of the first packet. We increase the packet counter by one with each incoming packet, regardless of its admission. When the counter reaches $T$, we set the counter to $0$ and set the values of $Min$ and $Max$ to the rank of the incoming packet (the resetting action).}
    \label{fig:rifo-resetting}
\end{figure*}

Due to carefully chosen admission criteria (designed for multi-criteria decision-making), this is sufficient for efficiency and robustness.
Our algorithm is built around the following three concepts:

\begin{enumerate}
\item
\textbf{Min and Max in a tracking range.} 
We maintain two statistics about the packets seen recently: the values $Min$ and $Max$.
We initialize them to the rank of the first packet we see and update them adequately as new packets come.
To ensure that the values characterize a recent range of packets, and further to avoid the effect of the outliers in classifying the ranks, we periodically reset them to the rank of the most recent packet; precisely, we do that every $T$ packets seen, where $T$ is a parameter, e.g. $T=100$.
The set of recent packets, called the \emph{tracking range}, starts as a single packet rank, and its length grows to the maximum size of $T$.
Fig.~\ref{fig:rifo-resetting} shows the resetting detail of \systemname.

\item
\textbf{Scoring by normalization.}
We score packets relatively to the ranks observed recently.
After determining the relative placement of the incoming packet's rank, we \emph{normalize} it to the recent ranks to obtain a score from the range $[0,1]$.
Inspired by Multi-Criteria Decision-Making~\cite{Triantaphyllou00,Tzeng11,JAHAN2015335}, for scoring a packet rank of rank $r_p$, we use a \emph{linear min-max normalization method}:

\begin{equation}
\label{eq:rifoRank}
    N(r_p) = \frac{r_p-Min}{Max-Min}.
\end{equation}

In the special case $Max = Min$, we cannot normalize according to the chosen method, and then admit the incoming packet.
    \item 
\textbf{Admission conditions.}
We admit the packet if either of the two conditions is satisfied:
\begin{enumerate}
    \item 
\textbf{Condition 1: comparing score with current queue length.}
We score the incoming packet's rank by a value in $[0,1]$ (the exact function defined above). The \ToN{lower} the score, the higher the chances for admission. We accept the packet whose score exceeds a threshold equal to the current utilization of the queue (0 for an empty queue, and 1 for a full queue).

 Precisely, \systemname checks the buffer size of the queue to admit the packets. Suppose that $B$ is the size of the buffer and $l$ is the number of packets in the buffer. We compute the proportion of buffer occupancy $\mathcal{Q}$ as follows:
 \begin{equation}
 \label{eq:bufferState}
     \mathcal{Q}= \frac{B-l}{B}
 \end{equation}

    \item 
\textbf{Condition 2: guaranteed admission buffer.} 
We leave a portion of a queue, called a \emph{guaranteed admission buffer} to serve the burst of traffic, and allow low-latency packet admission~\cite{aifo-sigcomm21}. 
A tunable parameter $k$ assigns a portion of queue size as a guaranteed admission buffer. 
If the current queue utilization is below $k$, we admit an incoming packet regardless of its rank.

\end{enumerate}

\end{enumerate}

\begin{algorithm}[t]
\SetAlgoNoLine
\SetKwInput{Initialization}{Initialization}
 \Initialization{}
 
\Indp
Let $B$ be the maximum length of the queue

Let $k$ be the fraction of the queue reserved for the guaranteed admission buffer

Let $T$ be the tracking range size

 Min := $\infty$
 
 Max := 0

 Counter := 0

\BlankLine


\Indm
\SetKwInOut{Ingress}{Ingress}
\Ingress{A packet of rank $r_p$}
\Indp

\If{Counter = T}{
    Min := Max := $r_p$ \Comment{reset Max and Min}

    Counter := 1
}
\Else{
    Min := $\min$($r_p$, Min)
    
    Max := $\max$($r_p$, Max)

    Counter := Counter+1
}

\BlankLine

\If{$Max = Min$}{
    Admit the packet
}
\Else{
Let $\ell$ be the current length of the queue
\If{$\ell \le k \cdot B$ or   $\frac{r_p - Min}{Max - Min} \le \frac{B - \ell}{B}$\color{black}}{
    Admit the packet
}
\Else{
    Drop the packet
}
}

\BlankLine

\Indm
\textbf{Egress:}

\Indp
\If{queue not empty}{
    Dequeue a packet and send
}

 \caption{The pseudocode of \systemname algorithm.}
 \label{alg:rifo}
\end{algorithm}

The pseudocode of \systemname is presented in Algorithm~\ref{alg:rifo}, and the architecture of the system is presented in Figure~\ref{fig:RIFO-architecture}. The main logic of the algorithm happens in the ingress part of the packet processing pipeline, where we compute the score of an incoming packet, compare it with the queue utilization, and decide if we admit the packet or drop it. At the egress, we dequeue packets and send them. Additionally, due to implementation limitations, we compute the queue length at egress and recirculate the packet with this information; for implementation details, we refer to \textsection\ref{sec:dataplane}.

\begin{figure}[tp]
    \centering
    \includegraphics[width=\linewidth]{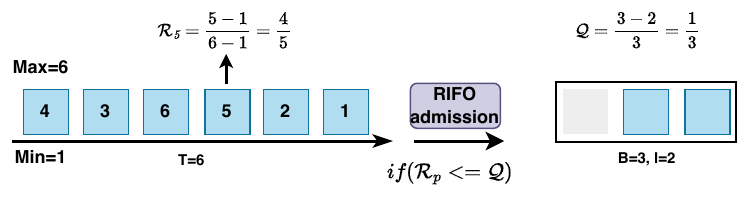}
    \caption{Example of \systemname admission with $T$=6 and $B$=3.}
    \label{fig:rifo-example}
\end{figure}

Let us now explain how \systemname admits the incoming packets into a FIFO queue. Consider an example scenario of Fig.~\ref{fig:rifo-example} with a tracking range size of six packets. In this tracking range, we have tracked six packets with $Min=1$ and $Max=6$. We assume there are already two packets in the queue ($l=2$) and now \systemname wants to decide for the packet with rank five. To this end, \systemname finds the relative rank of the packet using $Min$ and $Max$ values  ($\mathcal{R}_5=\sfrac{1}{5}$), and it checks the state of the buffer ($\mathcal{Q}=\sfrac{1}{3}$). Now, \systemname admits the packet if the relative rank condition check passes. Since $\mathcal{R}_5 \leq \mathcal{Q}$, it drops the packet. In this example, the relative rank of the packet with the rank of 5 --- on a range of 1 and 6 --- was not small enough to sacrifice the single empty slot in the queue to admit this packet. Only a packet of rank 4 (or lower) would have been admitted in this case.

\section{Analyzing the Accuracy of \systemname}\label{sec:analysis}

The admission criteria of \systemname estimates the quantile of the incoming packet in the distribution of the packet ranks in the traffic, and compares the quantile with the current buffer occupancy. Two design choices are made to keep the system simple and resource-efficient:
\begin{enumerate}
    \item 
    We approximate the quantile of incoming packets in the traffic's packet rank distribution and compare it with the current buffer occupancy.
    \item The packet rank distribution is represented by only the extreme values of recent packet ranks. These estimates are periodically reset.
\end{enumerate}

These design choices unavoidably lead to some loss of accuracy. However, they enable a straightforward and resource-efficient implementation that performs reasonably well in practice.
Ideally, we would like to maintain the exact distribution of the packet ranks, and compute the quantile of the incoming packet in this distribution.
This can however be computationally expensive, and require a large amount of memory.

Next, we delve into the impact of these design decisions on the system's accuracy compared to the ideal case. We first examine the quantile estimation accuracy assuming perfect Min and Max samples, followed by an analysis of the effect of the resetting period on the accuracy of the Min and Max samples.


\subsection{Quantile estimation by Min-Max normalization}

The approach of using mid-distribution values and piecewise linear functions to estimate quantiles of discrete distributions is a well-known technique, with formal guarantees for estimation quality established in prior research~\cite{Ma2011}. Although our method employs a single linear function, increasing the estimation error, the fundamental principle remains the same. Throughout this section, we assume a static underlying distribution.

If the underlying packet rank distribution is uniform, the normalization $\frac{r - Min}{Max - Min}$ gives us an \emph{exact} quantile estimation, because the CDF of a uniform distribution is $F(r) = \frac{r - Min}{Max - Min}$.
For a non-uniform distribution, the relationship between the normalization and the true quantile $F(r)$ is not straightforward. The expression $\frac{r - Min}{Max - Min}$ assumes a linear relationship between the rank and the quantiles, which is not the case for non-uniform distributions.

To analyze the error, let us denote the true quantile by $Q(r)$.
The error $E(r)$ in using the linear normalization as an estimate for the true quantile $Q(r)$ is $E(r)=Q(r)-\frac{r - Min}{Max - Min}$.
We can estimate this error by considering the first-order Taylor expansion near $r$ by $E(r) = (r-Min)f(Min) - \frac{r - Min}{Max - Min}$, where $f$ is the probability density function of the underlying distribution.

Informally, this error depends on how the underlying distribution deviates from the linear relationship assumed by the normalization. We can generally examine this deviation in terms of the mean $\mu$ and variance $\sigma^2$ of the distribution, to give intuitions how these values can affect the quality of our estimation.
\begin{enumerate}
    \item 
    Mean deviation. If the distribution is skewed such that the mean $\mu$ is not at the midpoint $\frac{Max+Min}{2}$, the linear mapping of the normalization will not capture this skewness. For instance, if $\mu > \frac{Max+Min}{2}$, the distribution has more weight on the higher end, meaning the normalization underestimates the true quantile for values near $Min$ and overestimates for values near $Max$.

    \item
    Variance deviation. Higher variance implies greater spread of the distribution, and if the distribution is not uniform, this affects the density differently across the range. Regions of higher density will have a steeper CDF, meaning the normalization will underestimate quantiles in high-density regions and overestimate in low-density regions.
\end{enumerate}

To upper bound the error, we can use the Kolmogorov-Smirnov statistic~\cite{daniel1990applied} (KS).
The KS statistic measures the discrepancy between the empirical distribution function of a sample and the cumulative distribution function of a reference distribution, or between the empirical distribution functions of two samples.
The KS statistic is defined as the supremum of the absolute difference between the empirical CDF and the true CDF $F(r)$ over all quantiles in the range $[Min, Max]$:
\[
    D = \sup_{r\in [Min,Max]} E(r) = \sum_{r\in[Min,Max]} |F(r) - \frac{r-Max}{Max-Min}|.
\]
For a given sample size $n$, the KS statistic is related to the sample maximum deviation $\epsilon_n$ between the empirical and true CDFs by the Dvoretzky-Kiefer-Wolfowitz (DKW) inequality~\cite{DvoretzkyKW}:
\[
    P(D>\epsilon_n) \le 2e^{-2n\epsilon_n^2}.
\]

The above bound gives us a method to estimate the error in the quantile estimation given that we can compute the discrepancy between the empirical and true CDFs for the sampled ranks. 

\color{black}

\subsection{Resetting the Min and Max samples}

\systemname keeps track of the minimum and maximum packet rank by maintaining two registers: Min and Max that are updated to the currently observed minimum and maximum packet rank, and reset periodically.
We investigate how closely these registers approximate the ground truth minimum and maximum across the entire sequence of packets.
Min and Max are the first and the last order statistics. Due to their sensitivity to outliers, these sample statistics can be reliably used only as estimators for light-tailed distributions or stationary processes~\cite{Haan.Ferreira2010}.

\begin{figure}[tp]
    \centering
    \includegraphics[width=0.8\linewidth]{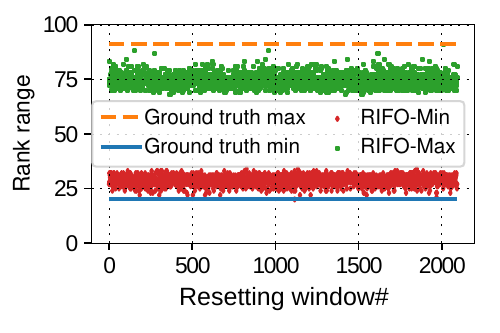}
    \caption{The influence of the resetting period $T$ on the Min and Max samples. The ground truth Min is 20, and the ground truth Max is 91. The sampled Min and Max values never deviate too much from the ground truth Min and Max, but they also do not reach the ground truth extremes of the underlying distribution.}
    \label{fig:resetting}
\end{figure}

In the extreme value theory, the Fisher-Tippett-Gnedenko theorem~\cite{fisher_tippett_1928} gives a characterization of the distribution of Min and Max statistics: depending on the source distribution, the extreme statistics can be distributed accordingly to either Fréchet distribution, Gumbel distribution or Weibull distribution, hence the asymptotic behavior of these samples is well-understood.
However, the Fisher-Tippett-Gnedenko theorem characterizes only the asymptotics of the Min and Max theorem --- as the resetting period $T$ tends to infinity.
Hence, for a fixed $T$, we empirically study the distribution of Min and Max statistics, revealing that for our data, even very small values of $T$ can track the ground truth of Min and Max statistics. We generate packets with uniform rank distribution between 0 and 100 in Netbench~\cite{netbench}. Fig.~\ref{fig:resetting} shows the influence of the resetting period, i.e., T=1000, on the Max and Min samples.

\section{Data Plane Design and Implementation}\label{sec:dataplane}

The \systemname admission algorithm is independent of any specific hardware architecture and can be used on Application-Specific Integrated Circuits (ASICs), Field Programmable Gate Array (FPGAs), and network processors. As an example, in this section, we state the data plane design of \systemname to implement on a modern programmable switch. This section aims to show the practicality of \systemname implementation in the aforementioned devices using P4~\cite{p4-ccr2014}. We implemented \systemname on a Tofino switch~\cite{tofino} using 650 lines of code in P4 language that can run at line rate. However, there are several challenges in implementing \systemname using P4 on Tofino, and herein, we describe how we overcome those challenges in more detail.
We note that Tofino switches were not designed to support floating point operations fully; however, recent work enables it~\cite{Yuan.NSDI22}.

\medskip
\noindent \textbf{Queue length estimation.}
\systemname needs to have information about queue lengths when the packet is being processed at the ingress pipe to decide about the admission by the traffic manager. However, this information is only available when the packet crosses the egress pipe, which is not accessible to the traffic manager anymore. More specifically, to admit the packets, we need to have the egress queue size in the traffic manager --- residing between the ingress and egress pipes. 

We obtain the egress queue occupancy information using a customized packet header for \systemname, which is called \textit{RIFO\_worker}. \systemname has $egress\_qlen$ as the header field to carry the egress queue length information on the recirculated packets. Then, \systemname recirculates the worker packet in the ingress pipe. When the recirculated worker packet reaches the egress pipe, we add the queue length information to the packet. Finally, the worker packets reach the ingress pipe of the device again.

\systemname uses the carried information by the worker packet in the traffic manager for evaluating Eq.~\ref{eq:bufferState}. Specifically, we need to compute $\frac{B-l}{B}$ to decide on admission. 



The worker packets of \systemname use the designated port for this purpose to read the egress queue length information. Therefore, these packets that are temporally generated by \systemname do not contribute to the queue occupancy of any specific egress port. 
If we assume that it takes 100ns to recirculate a packet on a port with \emph{10Mpps} rate, the mechanism will only add 1 extra packet to the network, which is negligible. We can also use one extra register to store the queue length information in the ingress pipe to reduce the overhead of recirculation in \systemname.
\systemname only needs the recirculation mechanism when the queue length is not yet available by the switch architecture. However, the second generation of Intel Barefoot Tofino switches~\cite{tofino} provides access to the queue length information in the ingress pipe, thus, removing the need for recirculation and its possible overhead and performance degradation.

\medskip
\noindent \textbf{\systemname admission condition.}
\systemname admits the packets based on their relative ranks, i.e., see Eq.~\ref{eq:rifoRank}, and queue occupancy, i.e., see Eq.~\ref{eq:bufferState}. Now, we explain how we compute the relative ranks of the packets in P4-enabled data planes since division and floating-point operations are not supported. We use approximation techniques to estimate the values of both equations~\cite{flexSwitch-nsdi17}. In doing so, we first apply the subtraction operations in Eq.~\ref{eq:rifoRank} to get the values of dividends and divisors.

We can use shift operation to approximate the values of both equations using the power of two numbers to simplify our data plane implementation. However, we can also transform the admission inequality into two multiplication operations. Specifically, we need to multiply the dividend value of Eq.~\ref{eq:rifoRank} to the capacity of the guaranteed admission buffer, i.e., $(1-k)\times B$, and the divisor of Eq.~\ref{eq:rifoRank} to the available buffer, i.e., $B-l$. The values of the guaranteed admission buffer and the available buffer may not be a power of two numbers. Therefore, we round these values to a power of two numbers. To further simplify the implementation of these operations, we use a lookup table with fixed values to obtain the result of each multiplication.
We transform the admission criteria Eq.~\ref{eq:rifoRank} and buffer occupancy  Eq.~\ref{eq:bufferState} to the following division-free form:
\begin{equation}
  (1-K) \times B \times (r_p - {Min}) \le (B - \ell) \times ({Max} - {Min})
\end{equation}
We can pre-compute the value \((1-K) \times B\) since the values of both parameters are known in advance. Thus, \systemname uses the exponent of the nearest power of two for this pre-computed value in the multiplication process. 

After this transformation, we need three lookup tables to calculate the exponent values of \((r_p - {Min})\), \(({Max} - {Min})\), and \((B - \ell)\).
Consider $\mathcal{V}$ be one of the values of $(r_p - Min)$, $ (Max - Min)$, and $ (B - \ell)$. We define the structure of the lookup tables for these values when they vary in the range of a 16-bit integer, as shown in Table~\ref{tab:val_hex}.
\begin{table}[htb]
\centering
\begin{tabular}{ccc}
\toprule
\textbf{$\mathcal{V}$} & \textbf{Mask} & \textbf{Exponent value} \\
\midrule
0x0002 & 0x0003 & 1 \\
0x0004 & 0x0007 & 2 \\
0x0008 & 0x000F & 3 \\
$\vdots$ & $\vdots$ & $\vdots$ \\
0x4000 & 0x7FFF & 15 \\
\bottomrule
\end{tabular}
\caption{Lookup table with to find the values of $(r_p - Min)$, $ (Max - Min)$, and $ (B - \ell)$ in approximating the admission criteria of \systemname.}
\label{tab:val_hex}
\end{table}

At this point, we need to perform two multiplication operations to get the values on the left and right sides of the inequality. We define two additional lookup tables to perform these multiplications. The first table uses the exponent value of \((1-K) \times B\) and \((r_p - {Min})\) as the keys, while the second table's keys are \(({Max} - {Min})\) and \((B - \ell)\). In both cases, we store the results of the multiplications as table entries. Finally, we retrieve the approximated values from these tables for comparison. \systemname admits the packets by comparing the values on the left and right sides of the transformed inequality.

\ToNMinor{
To illustrate how the lookup table and ternary matching simplify the calculations, consider an example where we want to compute the exponent value for \((r_p - {Min})\), one of the terms needed for the admission criteria. Suppose \((r_p - {Min})\) is \texttt{0x0008}. In our lookup table (Table~\ref{tab:val_hex}), we use a mask value in ternary matching to find the closest power of two. Specifically, for \texttt{0x0008}, we apply the mask \texttt{0x000F} to match this value to an exponent of 3, as it falls within the range specified by \texttt{0x0008 - 0x000F}. Thus, we retrieve an exponent value of 3 from the lookup table.
\\
For the multiplication, suppose \((1-K) \times B\) has a precomputed exponent of 4, while \((r_p - {Min})\) has an exponent of 3 obtained through the ternary match lookup. Using the first additional lookup table, where these exponents are the keys, we retrieve the result of \(2^{4+3} = 2^7\), calculating the multiplication as a power of two. This exponent-based approach efficiently computes multiplications without direct operations in the data plane.
\\
Following this approach, each lookup table uses a mask for ternary matching to convert values to their respective exponents, and then exponent combinations are used to approximate the multiplications required for the inequality comparison.
}

\medskip
\noindent {\textbf{Memory cost.} The memory cost associated with the lookup tables used for approximating the admission criteria of \systemname, assuming each parameter has a length of 16 bits, is summarized as follows: The lookup tables for approximating $(r_p - \text{Min})$, $(\text{Max} - \text{Min})$, and $(B - \ell)$ each have 15 entries occupies 36 bits. This results in a total memory usage of 67.5 bytes per table. The lookup table designed for calculating multiplication results based on exponent values contains 256 entries, each requiring 4 bytes, leading to a total memory usage of 1024 bytes. Therefore, the overall memory footprint is 1226.5 bytes. This compact memory requirement ensures that the proposed method is suitable for high-speed networking applications where memory efficiency is crucial.
\color{black}


\color{black}






\medskip
\noindent \textbf{Rank range resetting.}
\systemname keeps track of the number of arrival packets to reset the rank range values for $Min$ and $Max$ ranks registers.  We define the $rank\_range\_reg$ register for this purpose, and \systemname increments the value of this register by one unit by incoming each packet. When the value of this register reaches the predefined threshold value, i.e., $T$, to reset the $Min$ and $Max$ ranks registers, \systemname assigns the rank of the current packet as the  $Min$ and $Max$ values for both registers. The next round of tracking incoming packets starts after resetting the current value of the $rank\_range\_reg$ to zero.

\medskip
\noindent \textbf{RIFO header definition.}
\systemname uses the rank of the packets to decide on their admission and the end hosts generate them using a customized header. We add \systemname header on the top of UDP packets and use a predefined UDP destination port to distinguish \systemname packets at the switch. However, this is a design choice, and other protocols, such as TCP with the same destination port or other header layers, can also be used to distinguish the packets. The \systemname header contains only one field called \textit{rank}, and its size can be specified depending on the range of the ranks generated by the end hosts.

\section{Evaluation} \label{sec:eval}

To confirm the resource efficiency, we implemented \systemname on Tofino and studied resource consumption. To evaluate the performance of \systemname, we implemented a packet-level simulation and conducted extensive experiments under different workloads compared to state-of-the-art approaches. In this section, we report our main findings.
 
\begin{figure*}[tp]
    \centering
    \subfigure[Without \systemname\label{fig:tofino_udp_FIFO}]{\includegraphics[width=.4\linewidth]{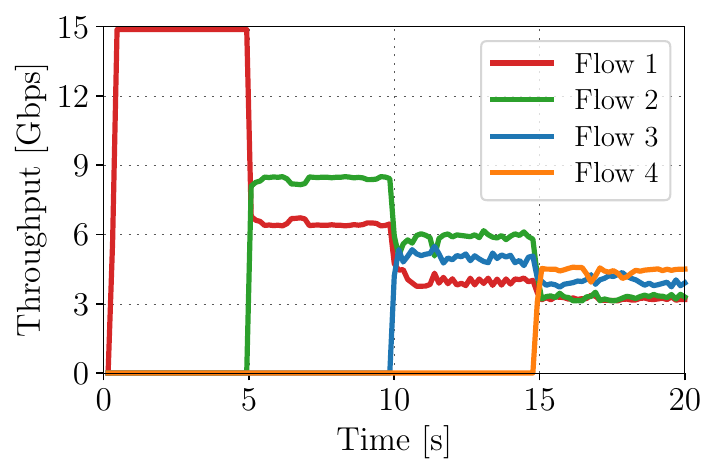}}
    \subfigure[RIFO\label{fig:tofino_udp_RIFO}]{\includegraphics[width=.4\linewidth]{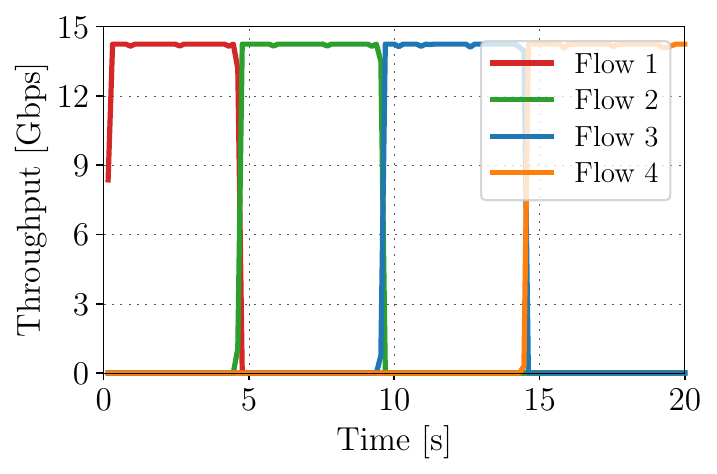}}
    \caption{Bandwidth split of without and with \systemname for scenarios with four flows when R(Flow 1)$~<$ R(Flow 2)$~<$ R(Flow 3)$~<$ R(Flow 4).}
    \label{fig:tofino}
\end{figure*}

\subsection{Hardware Testbed}


\noindent\textbf{Low resource consumption and its implications.}
The proposed scheduler is resource-efficient, leaving the crucial resources of the switch for other purposes. Many network functions are memory- and CPU- hungry; for example, separate queues are maintained for each tenant to enable multitenancy. More memory helps more accurate measurements of network streams; for example, the error of count sketches rapidly increases when the memory is insufficient. Memory is also used in stateful I/O and machine learning on edge.

To emphasize resource efficiency, we report the amount of the resource consumption by \systemname and compare it with state-of-the-art systems such as AIFO and SP-PIFO on a Tofino switch implemented using \emph{bf-sde-9.11.0}. \systemname and AIFO use one FIFO switch to schedule the packets, while SP-PIFO does it with multiple strict priority queues. Additionally, AIFO and SP-PIFO need to store multiple states to admit the packets. More specifically, AIFO uses registers to keep the latest sampled packets, and SP-PIFO leverages them to maintain queue bounds. Table~\ref{tab:resource-consumption} shows that \systemname leverages significantly fewer resources than AIFO and SP-PIFO in all different resource types. For instance,  \systemname needs 2.54x and 6.55x less SRAM compared with AIFO and SP-PIFO, respectively.

\begin{table}[tp]
    \centering
\begin{tabular}{lccc}
\toprule
\textbf{Resource type} & \textbf{\systemname} & \textbf{AIFO} & \textbf{SP-PIFO}  \\
\midrule
Match Crossbars & 2.57\% & 10.74\% & 8.05\% \\
\midrule
Gateway & 5.36\% & 3.2\% & 10.62\%\\
\midrule
Hash Bits & 2.27\% & 1.92\% & 4.81\%\\
\midrule
SRAM & 2.86\% & 7.29\% & 18.75\%\\
\midrule
TCAM & 1.79\% & 0\% & 0.42\%\\
\midrule
Stateful ALUs & 14.29\% & 47.92\% & 20\%\\
\midrule
Logical Table IDs & 17.86\%  & 26.56\% & 18.12\%\\
\bottomrule
\\
\end{tabular}
\caption{The resource consumption of \systemname, and reference implementations of AIFO and SP-PIFO on Intel Barefoot Tofino. The values are presented in percentage.} \label{tab:resource-consumption}
\end{table}

\noindent\textbf{Bandwidth split.}\ \
We conduct the performance evaluation of \systemname on our hardware testbed, which comprises a \textit{Netberg Aurora 710} Tofino switch with a capacity of 3.2Tbps~\cite{netberg}. The switch connects two servers via \textit{Mellanox ConnectX-5} adapters utilizing 100Gbps links. The sender side operates on an 8-core Intel(R) Core(TM) i7-6900K CPU @ 3.20GHz, while the receiver side utilizes an Intel(R) Xeon(R) w7-3465X CPU. Both sender and receiver modules are developed using DPDK version 22.11.1~\cite{dpdk}. The servers run \textit{Ubuntu 22.04.3 LTS} operating system, with the Linux kernel version \textit{6.2.0-39-generic}. 


\ToNMinor{To demonstrate the prioritization capabilities of \systemname, we throttle the link capacity from the switch to the receiver to 15 Gbps, creating a congested scenario. This configuration enables us to examine how the system prioritizes high-priority packets under network constraints. While the hardware supports 100Gbps, the throttling allows us to simulate contention and evaluate \systemname’s handling of competing flows at various priorities, which would not be as evident under uncongested 100Gbps conditions.}

We conduct bandwidth split tests without and with \systemname with four UDP flows, each at 15Gbps, over a bottleneck link. These flows have different priorities, as illustrated in previous works~\cite{sp-pifo-nsdi20,aifo-sigcomm21,hcsfg-nsdi21}. We initiate one flow every five seconds from the sender, where the packets of \textit{Flow 4} hold the lowest rank (highest priority), while those of \textit{Flow 1} hold the highest rank. With four concurrent flows, i.e., \textit{Flow 1} to \textit{Flow 4}, each with 15Gbps, we generate 60Gbps traffic from the source to the destination, thus creating a congested link. 

Figure~\ref{fig:tofino} illustrates how the switch allocates bandwidth to these flows without and with  \systemname. Without \systemname tries to evenly distribute bandwidth upon the arrival of each new flow every five seconds, ensuring equitable sharing among active flows, as depicted in Figure~\ref{fig:tofino_udp_FIFO}. In contrast, \systemname dynamically allocates available bandwidth based on the rank of the flows, with the highest-ranking flow monopolizing the link's capacity in Figure~\ref{fig:tofino_udp_RIFO}. For instance, during the second five-second interval, \systemname assigns the bottleneck link capacity to the packets of \textit{Flow 2}, which hold the highest rank, while the packets of \textit{Flow 1} experience drops.

\subsection{Packet-level Simulation}\label{sec:NetbenchResults}

We perform packet-level simulation similar to the state-of-the-art packet scheduling algorithms~\cite{aifo-sigcomm21,sp-pifo-nsdi20,pifo-sigcomm16} on a leaf-spine datacenter topology with nine leaf switches, four spine switches, and 144 servers. \ToN{ The bandwidth of access links is 100 Gbps, while the leaf-spine links are 400 Gbps.} We implement \systemname in Netbench~\cite{netbench}.

\begin{figure*}[ht]
    \centering
    \subfigure[Average FCT of small flows\label{fig:avgFCT-websearch}]{\includegraphics[width=.32\linewidth]{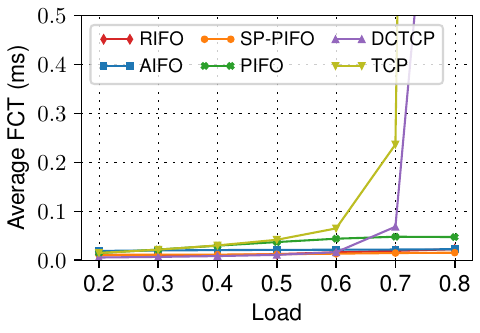}}
    \subfigure[99th FCT of small flows\label{fig:FCT99-websearch}]{\includegraphics[width=.32\linewidth]{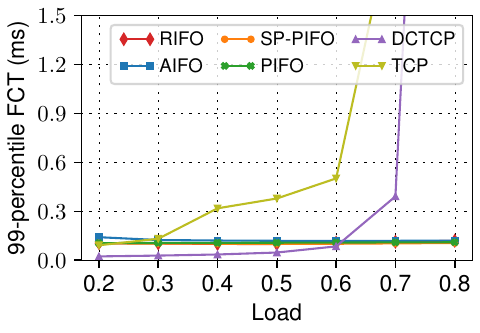}}
    \subfigure[Average FCT of large flows\label{fig:avgFCTLarge-websearch}]{\includegraphics[width=.32\linewidth]{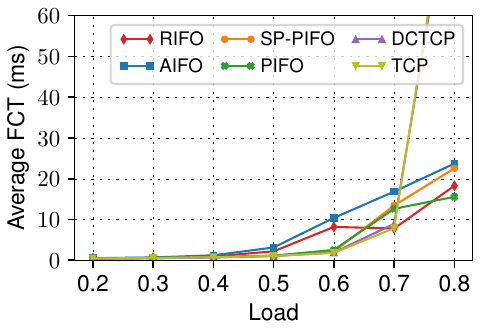}}
    \caption{Simulation results on websearch workload for minimizing FCT.}
    \label{fig:FCT-WS}
\end{figure*}

\begin{figure*}[ht]
    \centering
    \subfigure[Average FCT of small flows\label{fig:avgFCT-dataMining}]{\includegraphics[width=.32\linewidth]{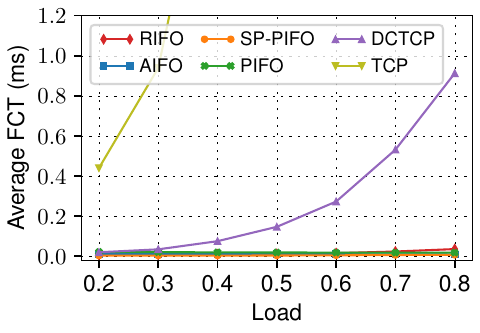}}
    \subfigure[99th FCT of small flows\label{fig:FCT99-dataMining}]{\includegraphics[width=.32\linewidth]{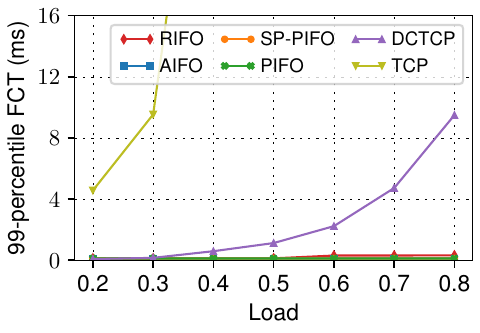}}
    \subfigure[Average FCT of large flows\label{fig:avgFCTLarge-dataMining}]{\includegraphics[width=.32\linewidth]{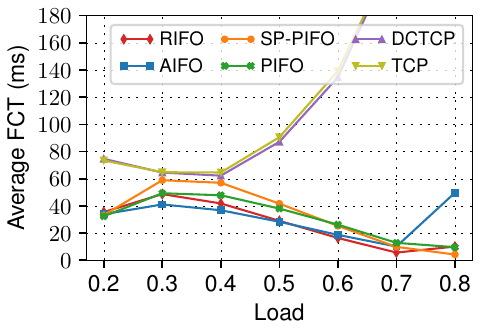}}
    \caption{Simulation results on datamining workload for minimizing FCT.}
    \label{fig:FCT-dm}
\end{figure*}

We evaluate the performance of programmable packet schedulers for two objectives: minimizing FCT and fairness. 
We compare the performance of pFabric when run on top of \systemname, PIFO~\cite{pifo-sigcomm16}, SP-PIFO~\cite{sp-pifo-nsdi20}, and AIFO~\cite{aifo-sigcomm21}. We also report the FCT flows for TCP NewReno with drop-tail and DCTCP with ECN-marked drop-tail queues. We run the experiments on commonly used by the networking community in the datacenter scenarios, i.e., websearch~\cite{pFabric-sigcomm13} and datamining~\cite{datamininWorkload-sigcomm09}.
We also conduct experiments to check the effect of tracking range and queue sizes on the performance of \systemname.
\ToN{We set the initial values of the target queue length to 20 ($B=20$), the track range size to 500 ($T=500$), and use 10\% of the buffer as the guaranteed admission buffer similar to AIFO~\cite{aifo-sigcomm21}.}

\subsubsection{FCT Minimization with \systemname}
In this section, we report how \systemname performs in minimizing FCT under realistic workloads, i.e., websearch and datamining. We implement Shortest Remaining Processing Time (SRPT) for pFabric~\cite{pFabric-sigcomm13} to compute the rank of incoming packets and assess the performance of \systemname. We generate the traffic load according to the Poisson distribution of the workloads and report the results when the load varies in the range of 0.2 to 0.8. We classify the generated flows into small ($<$~100KB) and large ($\geq$~1MB) ones in our evaluation. Our evaluations also include traffic generated according to Pareto distribution shedding new light on the performance of AIFO.

We compare the performance of \systemname in terms of FCT minimization with those of SP-PIFO, PIFO, and AIFO, which are the state-of-the-art packet scheduling algorithms in the data plane as well as TCP and DCTCP. Furthermore, pFabric is the transport layer at the end-hosts for the programmable packet schedulers.

\ToN{Fig.~\ref{fig:FCT-WS} presents the FCT of different flows for websearch workload. More specifically, Fig.~\ref{fig:avgFCT-websearch} shows all the programmable packet schedulers, i.e., \systemname, AIFO, SP-PIFO, and PIFO, achieve a similar average FCT for small flows regardless of the load. However, the average FCT of small flows for PIFO slightly increases when the network has a higher load. Fig~\ref{fig:FCT99-websearch} shows that all programmable packet schedulers achieve similar 99th percentile FCT for the small flows. Fig.~\ref{fig:avgFCTLarge-websearch} presents the average FCT of large flows for the same workload. \systemname reduces the FCT of large flows compared with AIFO when the load is higher than 0.4, and this FCT improvement happens compared with SP-PIFO when the load is higher than 0.6. We note that \systemname and AIFO use just one single FIFO queue.
We also observe that the average FCT of large flows increases when we inject more traffic into the network. One reason for such an FCT increment is that the packet schedulers prefer transmitting the packets of small flows, and packet drops from large flows are inevitable.}


\ToN{We now repeat the same experiments for the datamining workload, and Fig.~\ref{fig:FCT-dm} reports the FCT of different flow sizes. Fig.~\ref{fig:avgFCT-dataMining} shows that the small flows achieve a similar average FCT when running with programmable packet schedulers. However, the 99-percentile FCT of small flows in \systemname is slightly higher than others when the network load is higher than 0.5 in Fig.~\ref{fig:FCT99-dataMining}.  
TCP has the highest FCT for small flows among all algorithms. For large flows, \systemname can achieve lower FCT up to 2.25x, 1.73x, and 4.91x compared with PIFO, SP-PIFO, and AIFO (see Fig.~\ref{fig:avgFCTLarge-dataMining}). The main reason behind such improvement relies on the admission mechanism of \systemname with fixed tracking range size, which is beneficial for scenarios with larger flows. Due to the large number of small flows under high loads, the average FCT of large flows benefits this proportionally. }



\begin{figure*}[ht]
    \centering
    \subfigure[Average FCT of small flows\label{fig:avgFCT-websearch-varyRS}]{\includegraphics[width=.32\linewidth]{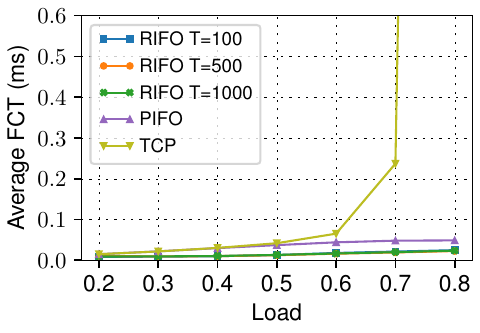}}
    \subfigure[99th FCT of small flows\label{fig:FCT99-websearch-varyRS}]{\includegraphics[width=.32\linewidth]{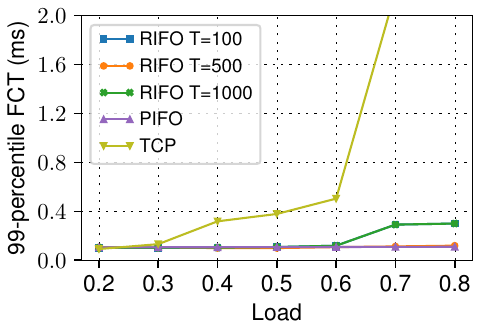}}
    \subfigure[Average FCT of large flows\label{fig:avgFCTLarge-websearch-varyRS}]{\includegraphics[width=.32\linewidth]{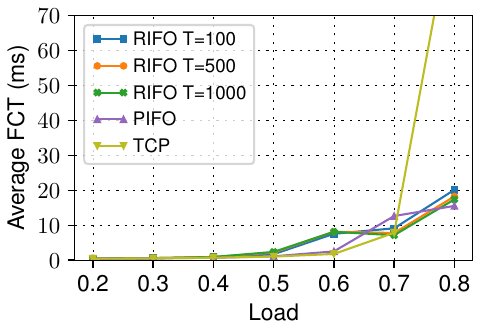}}
    \caption{The effect of tracking range size on the FCT of different flow sizes of \systemname.}
    \label{fig:FCT-WS-varyRS}
\end{figure*}

\subsubsection{Impact of the tracking range size}\label{sec:tarck-range-eval}
We now study the impact of tracking range size on the FCT of different flows for the websearch workload. 
\ToN{This parameter controls the values for the $Min$ and $Max$ parameters, which play the main role in dropping packets. For the first set of experiments in Fig.~\ref{fig:FCT-WS} and Fig.~\ref{fig:FCT-dm}, we set the tracking range size to 500 packets. 
\\
Fig.~\ref{fig:FCT-WS-varyRS} shows the results of experiments when the track range is 100, 500, and 1000 packets. We observe that for scenarios when the link load is less than 50\%, increasing the track range has minimal impact on the FCT of flows regardless of their sizes. }
We note that a smaller track range improves the FCT of small flows, while large values are good for large flows. We summarize the main reason for such an impact as follows. When the track range is small, we have fewer packet ranks to find the relative rank of each packet. This condition is advantageous for small flows since the probability of having large fluctuations among the packets of different flows decreases. In the case of a large track range, we may admit packet ranks that belong to a very large flow, and the probability of having another set of such flows is less according to the traffic flow size distribution of websearch. Therefore, \systemname is more efficient in admitting the packets of large flows.

\subsubsection{Impact of Queue Size}
\ToN{We measure the impact of queue length on the FCT of small and large flows obtained by \systemname when running the simulations using the websearch workload. We report how different queue lengths affect the FCT and the throughput of flows.
We first set the $B$ to a small value like 10 packets, and then increase it to 500.
\\
Fig.~\ref{fig:avgFCT-small-websearch-qLen} shows that \systemname achieves a similar FCT for small flows regardless of the queue length when the load is less than 60\%. However, for scenarios with a load higher than 60\%, adding more space to the queue results in the FCT increment for small flows since \systemname admits more small flows with similar relative ranks. Accumulating more small flows in the single FIFO queue leveraged by the \systemname results in longer FCTs. However, we observe the opposite behavior for the same experiments but for large flows in Fig~\ref{fig:avgFCT-large-websearch-qLen}. Having a larger queue length is desired when the network load is high since it results in FCT reduction for them.
\\
Fig.~\ref{fig:avgThroughput-all-websearch-qLen} shows that flows achieve higher throughput when the network has a lower load and higher queue length. The main reason for such results is that in such scenarios, \systemname can admit more packets from different flows, and the likelihood of dropping packets is lower. }

\begin{figure*}[tp]
    \centering
    \subfigure[Average FCT of small flows\label{fig:avgFCT-small-websearch-qLen}]{\includegraphics[width=.32\linewidth]{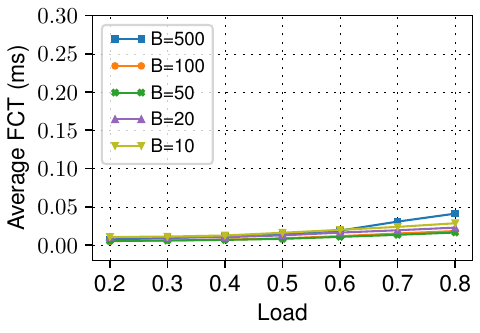}}
    \subfigure[Average FCT of large flows\label{fig:avgFCT-large-websearch-qLen}]{\includegraphics[width=.32\linewidth]{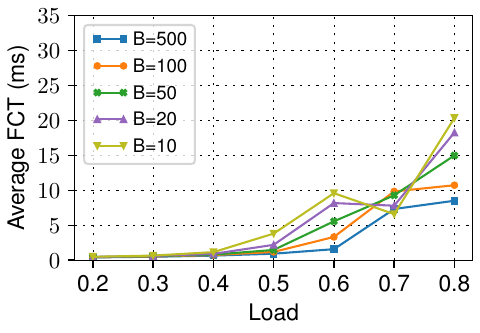}}
    \subfigure[Average throughput for all flows\label{fig:avgThroughput-all-websearch-qLen}]{\includegraphics[width=.32\linewidth]{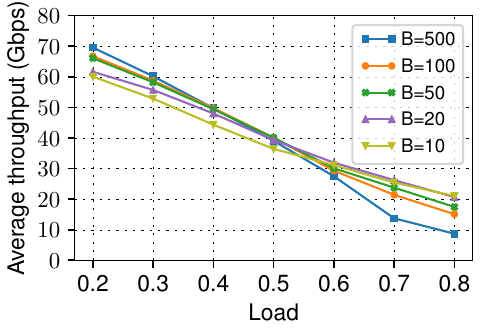}}
    \caption{The effect of queue size on the FCT and throughput of flows running \systemname.}
    \label{fig:FCT-WS-qLen}
\end{figure*}

\subsubsection{Impact of T and B}\label{sec:tarck-range-eva}
In this section, we study the impact of a large tracking range $T$ for scenarios with a larger queue size $B$. Since we already reported the impact of $T$ and $B$ when one of them is constant, we now equate the values of both parameters and report the impact on the FCT and throughput. The choice of equal parameters $T=B$ is especially attractive, as it relieves network operators from the necessity of setting the parameter $T$. 

\begin{table}[tp]
    \centering
\begin{tabular}{lccc ccc}
\toprule
& \multicolumn{3}{c}{RIFO} &  \multicolumn{3}{c}{AIFO} \\\cmidrule(l{6pt}r{6pt}){2-4}  \cmidrule(l{6pt}r{6pt}){5-7}
\textbf{Load} &\textbf{ B=50} &  \textbf{B=100} & \textbf{ B=500} &  \textbf{B=50} &  \textbf{B=100} &  \textbf{B=500} \\
\midrule
\textbf{0.2} & 0.005 & 0.005 & 0.007 & 0.013 & 0.010 & 0.009 \\
\textbf{0.3} & 0.005 & 0.006 & 0.009 & 0.015 & 0.012 & 0.012 \\
\textbf{0.4} & 0.006 & 0.007 & 0.011 & 0.017 & 0.014 & 0.016 \\
\textbf{0.5} & 0.008 & 0.008 & 0.014 & 0.018 & 0.015 & 0.020 \\
\textbf{0.6} & 0.010 & 0.011 & 0.018 & 0.018 & 0.015 & 0.024 \\
\textbf{0.7} & 0.013 & 0.015 & 0.030 & 0.018 & 0.014 & 0.030 \\
\textbf{0.8} & 0.016 & 0.018 & 0.041 & 0.018 & 0.013 & 0.034 \\
\bottomrule
\\
\end{tabular}
\caption{FCT of small flows for \systemname and AIFO when T=B.} \label{tab:FCT-T=B-Small}
\end{table}

\begin{table}[tp]
\begin{tabular}{lccc ccc}
\toprule
& \multicolumn{3}{c}{RIFO} &  \multicolumn{3}{c}{AIFO} \\\cmidrule(l{6pt}r{6pt}){2-4}  \cmidrule(l{6pt}r{6pt}){5-7}
\textbf{Load} &\textbf{ B=50} &  \textbf{B=100} & \textbf{ B=500} &  \textbf{B=50} &  \textbf{B=100} &  \textbf{B=500} \\
\midrule
\textbf{0.2	} &0.45	&0.44	&0.41&	0.48	&0.47&	0.43\\
\textbf{0.3} & 0.58 & 0.55 & 0.51 & 0.68 & 0.63 & 0.54 \\
\textbf{0.4} & 0.82 & 0.75 & 0.67 & 1.14 & 1.00 & 0.71 \\
\textbf{0.5} & 1.48 & 1.20 & 0.94 & 2.65 & 2.05 & 1.05 \\
\textbf{0.6} & 5.57 & 3.34 & 1.59 & 8.08 & 6.27 & 2.03 \\
\textbf{0.7} & 9.29 & 9.85 & 7.34 & 15.18 & 14.29 & 10.02 \\
\textbf{0.8} & 15.00 & 10.74 & 8.52 & 20.52 & 18.77 & 14.38 \\
\bottomrule
\\
\end{tabular}
\caption{FCT of large flows for \systemname and AIFO when T=B.} \label{tab:FCT-T=B-Large}
\end{table}

\ToN{
Table~\ref{tab:FCT-T=B-Small} and Table~\ref{tab:FCT-T=B-Large} show that the FCT of flows in \systemname and AIFO improves when we use equal value for $T$ and $B$ for both small and large flows. \systemname achieves approximately $3$x shorter FCT for small flows, specifically when the network has a lower load, while both approaches achieve a similar FCT for small flows when the B is 500. 
\systemname gains high throughput when the T and B are higher, and the network load is low. However, increasing the load has the opposite impact on the average FCT of all flows regardless of the values of T and B. 
We also note that increasing the size of the guaranteed admission buffer (the parameter~$k$) harms the FCT of the small flows.
}

\begin{figure}[ht]
    \centering
    \subfigure[websearch workload\label{fig:Throughput-all-webSearch}]{\includegraphics[width=.48\linewidth]{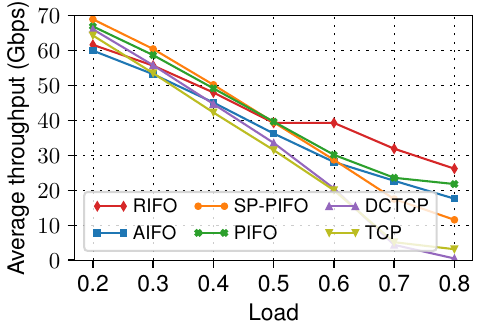}}
    \subfigure[datamining workload\label{fig:Throughput-all-dataMining}]{\includegraphics[width=.48\linewidth]{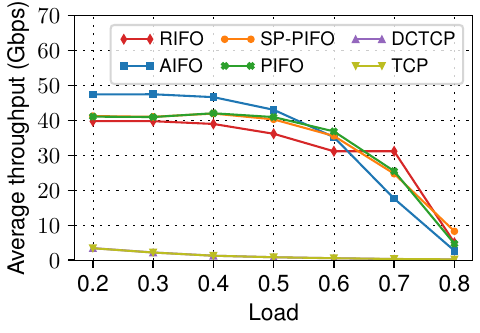}}

    \caption{The average throughput of large flows for various packet schedulers.}
    \label{fig:throughput}
\end{figure}

\subsubsection{Throughput}
In this section, we report the average throughput of large flows for websearch (see Fig.~\ref{fig:Throughput-all-webSearch}) and data-mining workloads (see Fig.~\ref{fig:Throughput-all-dataMining}).
We calculate the throughput of the large flows as the ratio between the amount of all the received bytes and the sum of the FCTs.
We observe the highest throughput when the load is 0.2, and increasing the load results in dropping throughput regardless of any packet scheduling algorithms in both workloads. We also observe that depending on the load, the average throughput of the schedulers changes significantly. For example, the websearch workload contains more small flows than those of datamining, and consequently, the throughput of large flows is lower. Across the board, \systemname gains higher throughput than AIFO in most link loads even though both use a FIFO queue for admission. SP-PIFO has better throughput than \systemname and AIFO since it leverages multiple strict priority queues.

\subsubsection{Fair Queueing}
In this section, we study the impact of fair queueing on the performance of packet scheduling algorithms. We implement fair queueing for flows using Start-Time Fair Queueing
(STFQ)~\cite{stfq-sigcomm96} to assign ranks on top of AIFO, SP-PIFO, and PIFO. We also include the state-of-the-art fair queueing mechanisms such as TCP, DCTCP, and AFQ~\cite{afq-nsdi18} to compare the fairness. Fig.~\ref{fig:FCT-WS-wfq} reports the FCT of different flows on websearch workload. Specifically, Fig.~\ref{fig:avgFCT-websearch-wfq} shows the average FCT of small flows while the 99th percentile latency of small flows was shown in Fig.~\ref{fig:FCT99-websearch-wfq}. We also report the FCT of large flow sizes in Fig.~\ref{fig:avgFCTLarge-websearch-wfq}. We note that in most scenarios, \systemname achieves similar performance to those of AIFO, SP-PIFO, and PIFO by using a single tracking range and a FIFO queue.

\begin{figure*}[tp]
    \centering
    \subfigure[Average FCT of small flows\label{fig:avgFCT-websearch-wfq}]{\includegraphics[width=.32\linewidth]{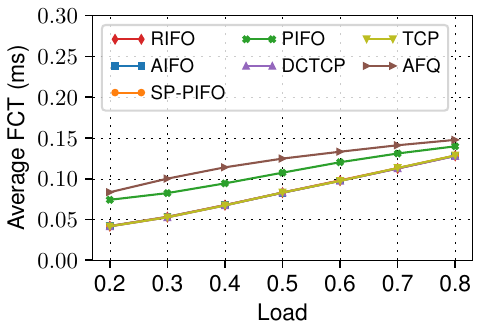}}
    \subfigure[99th FCT of small flows\label{fig:FCT99-websearch-wfq}]{\includegraphics[width=.32\linewidth]{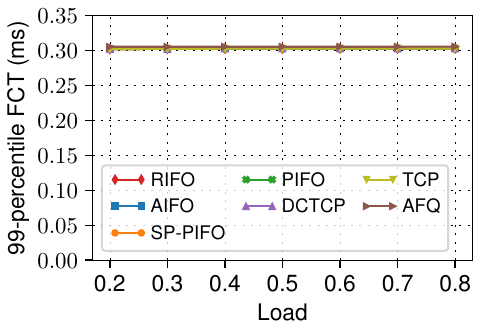}}
    \subfigure[FCT of large flows\label{fig:avgFCTLarge-websearch-wfq}]{\includegraphics[width=.32\linewidth]{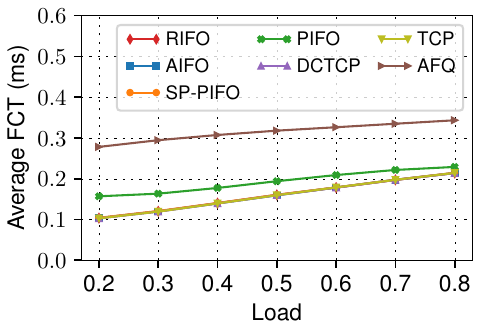}}
    \caption{The impact of fair queuing on the FCT of websearch workload.}
    \label{fig:FCT-WS-wfq}
\end{figure*}

We now report the impact of fair queueing on the FCT of flows for datamining workload in Fig.~\ref{fig:FCT-DM-wfq}. We can observe similar trends in the performance of all algorithms to those of websearch, including for the datamining workload. However, we have shorter FCTs for small flows, while large flows have longer FCTs.

\begin{figure*}[tp]
    \centering
    \subfigure[Average FCT of small flows\label{fig:avgFCT-datamining-wfq}]{\includegraphics[width=.32\linewidth]{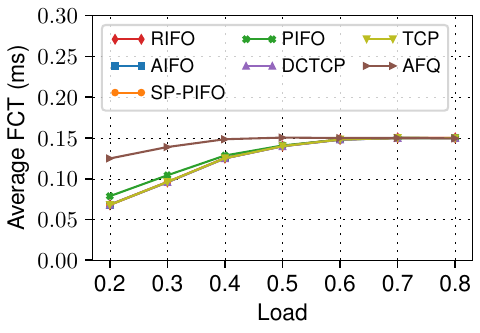}}
    \subfigure[99th FCT of small flows\label{fig:FCT99-datamining-wfq}]{\includegraphics[width=.32\linewidth]{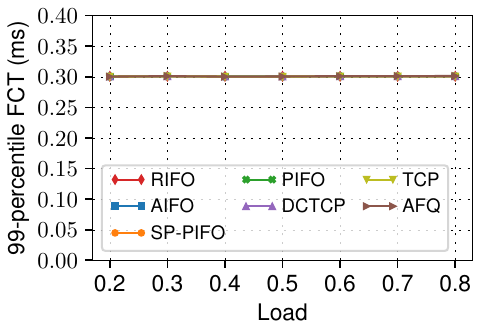}}
    \subfigure[FCT of large flows\label{fig:avgFCTLarge-datamining-wfq}]{\includegraphics[width=.32\linewidth]{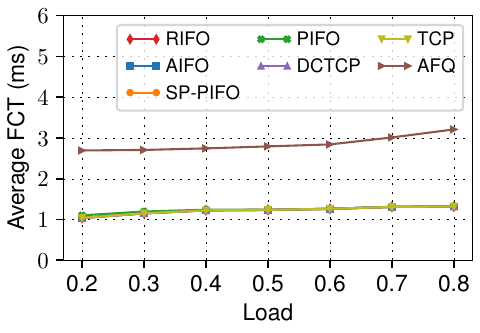}}
    \caption{The impact of fair queuing on the FCT of datamining workload.}
    \label{fig:FCT-DM-wfq}
\end{figure*}

\subsubsection{Experiments Using Pareto Distribution.}
In this section, we measure the impact of using Pareto distribution for incoming traffic. We generate the traffic load according to the Pareto distribution of the workloads and report example results when the load varies from 0.2 to 0.8. We set the Pareto distribution \textit{shape} parameter value to 1.05 and use the Netbench~\cite{netbench} implementation to find the corresponding value for the scale parameter. Fig.~\ref{fig:FCT-WS-pareto} shows the FCT of different algorithms on websearch workload. Across the board, DCTCP has the performance in all experiments, while RIFO and SP-PIFO have very similar FCTs for various flow sizes. 

Previous evaluations of AIFO reported superior performance compared to SP-PIFO, whereas our new experiments for Pareto distribution demonstrate that SP-PIFO performs better than AIFO. Among the programmable packet schedulers, RIFO and SP-PIFO achieve the lowest FCTs for small flows while their performance is similar for large flows.

\begin{figure*}[tp]
    \centering
    \subfigure[Average FCT of small flows\label{fig:avgFCT-websearch-pareto}]{\includegraphics[width=.32\linewidth]{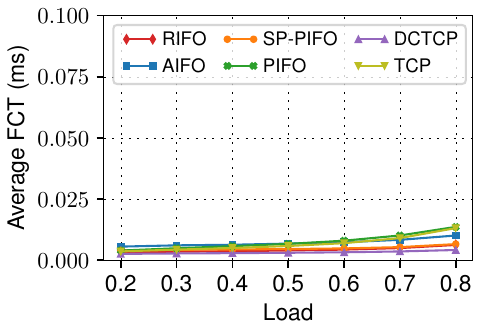}}
    \subfigure[99th FCT of small flows\label{fig:FCT99-websearch-pareto}]{\includegraphics[width=.32\linewidth]{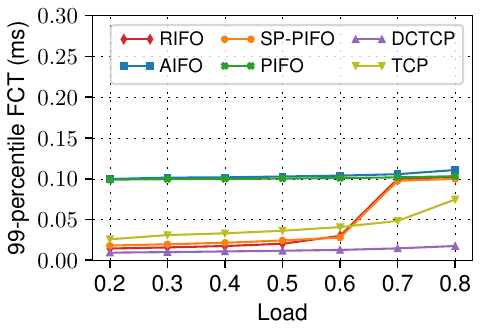}}
    \subfigure[FCT of large flows\label{fig:avgFCTLarge-websearch-pareto}]{\includegraphics[width=.32\linewidth]{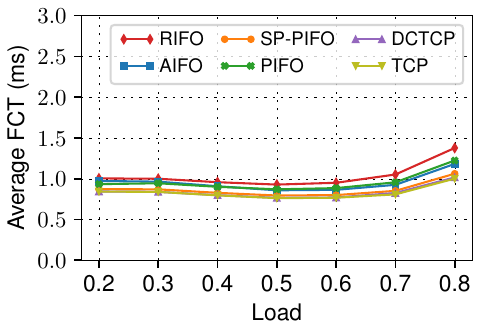}}
    \caption{The impact of Pareto distribution on the FCT of websearch workload.}
    \label{fig:FCT-WS-pareto}
\end{figure*}

\section{Related Work}\label{sec:relatedWork}

Programmable packet scheduling has recently received much attention~\cite{TowardsProgrammable,pifo-sigcomm16}
and is an example of a more general trend pushing innovation in communication networks
using programmable data planes~\cite{csur21}.
Programmable packet scheduling approaches can closely approximate traditional, fixed-function schedulers such as Shortest Remaining Processing Time~\cite{srpt-OR66} for flow completion time and Start-Time Fair Queuing~\cite{stfq-sigcomm96} for fairness, and can express others such as Least-Slack-Time-First~\cite{Leung2005ANA} and 
 Service-Curve Earliest Deadline First~\cite{SCED}.
In addition to realizing optimization objectives, it is also possible to account for other requirements such as privacy~\cite{IFS}.
Despite its flexibility, it was argued that programmable scheduling cannot jointly optimize all the desired objectives~\cite{UniversalPacketScheduling} (an observation following a similar result for traditional schedulers~\cite{SilverBullet}).

Push-In-First-Out (PIFO)~\cite{pifo-sigcomm16} is often considered an ideal programmable scheduling algorithm. 
However, it is complex and hence, over the last years, several algorithms for programmable packet scheduling were proposed that approximate only some aspects of PIFO, but are compatible with available programmable switches at line rate at scale.
AIFO~\cite{aifo-sigcomm21}, an admission-only approximation of PIFO with a single FIFO queue was thoroughly discussed across the paper. We note that AIFO can be implemented with less memory by decreasing the tracking window size, but no reference implementation exists.
SP-PIFO~\cite{sp-pifo-nsdi20} maintains a sequence of FIFO queues, assigns the packet to one of them depending on its rank, and dequeues the packets from the head.
Each queue maintains its \emph{threshold} for admission: a packet is assigned to the first queue with a threshold lower than the packet rank.
The thresholds dynamically adjust to the ranks:
(1) to minimize inversions within one queue, upon admission, the threshold is at least the rank of the admitted packet, and (2) if the inversion in the last queue is detected, all thresholds decrease.

To drive the development of programmable switch hardware, several other systems for programmable scheduling were proposed~\cite{Eiffel, AppxFQ, FastHardware}\ToN{\cite{Shrivastav19,AtreSS24,YaoZFGLFXC23}} that rely on new hardware designs.

There is also interesting work on possible extensions of the rank-based programmable scheduling approach. In particular,
calendar queues \cite{calendarQueues-NSDI20,hcsfg-nsdi21} generalize the rank-based scheduling approach, and enable more sophisticated packet prioritization policies, where scheduling decisions depend not only on the packet rank but also on the elapsed time. Gearbox~\cite{gearbox-nsdi22} employs multiple FIFO queues to approximate WFQ to allocate bandwidth.
\section{Conclusion and Future Directions}\label{sec:conclusion}
We studied the efficiency of packet scheduling in the programmable data plane and proposed 
\systemname, a simple yet effective algorithm that uses only three mutable registers and one FIFO queue, and is implementable on existing hardware at line rate.

We see our work as a first step and believe that it opens several interesting directions for future research.
In particular, it will be interesting to explore alternative normalization methods, some of which may become available 
only as the capacities of programmable switches grow (such as enabling floating-point operations~\cite{Yuan.NSDI22}). Indeed, various normalization methods exist in multi-criteria decision-making, and we demonstrated the power of one of the simplest and most basic methods.
It would also be interesting to conduct a more in-depth study of the effect of different switch resources on performance and energy consumption.

\balance





\section*{Acknowledgment}
In memory of Alessandro di Bucchianico for his feedback on analyzing min-max normalization to approximate quantile computation.
This work was partially funded by the Austrian Science Fund (FWF), project I 5025-N (DELTA), and by the German Research Foundation (DFG), SPP 2378 (ReNO), 2023-2027.

\bibliographystyle{IEEEtran}
\bibliography{paper}

\begin{IEEEbiography}[{\includegraphics[width=1in,height=1.25in,clip,keepaspectratio]{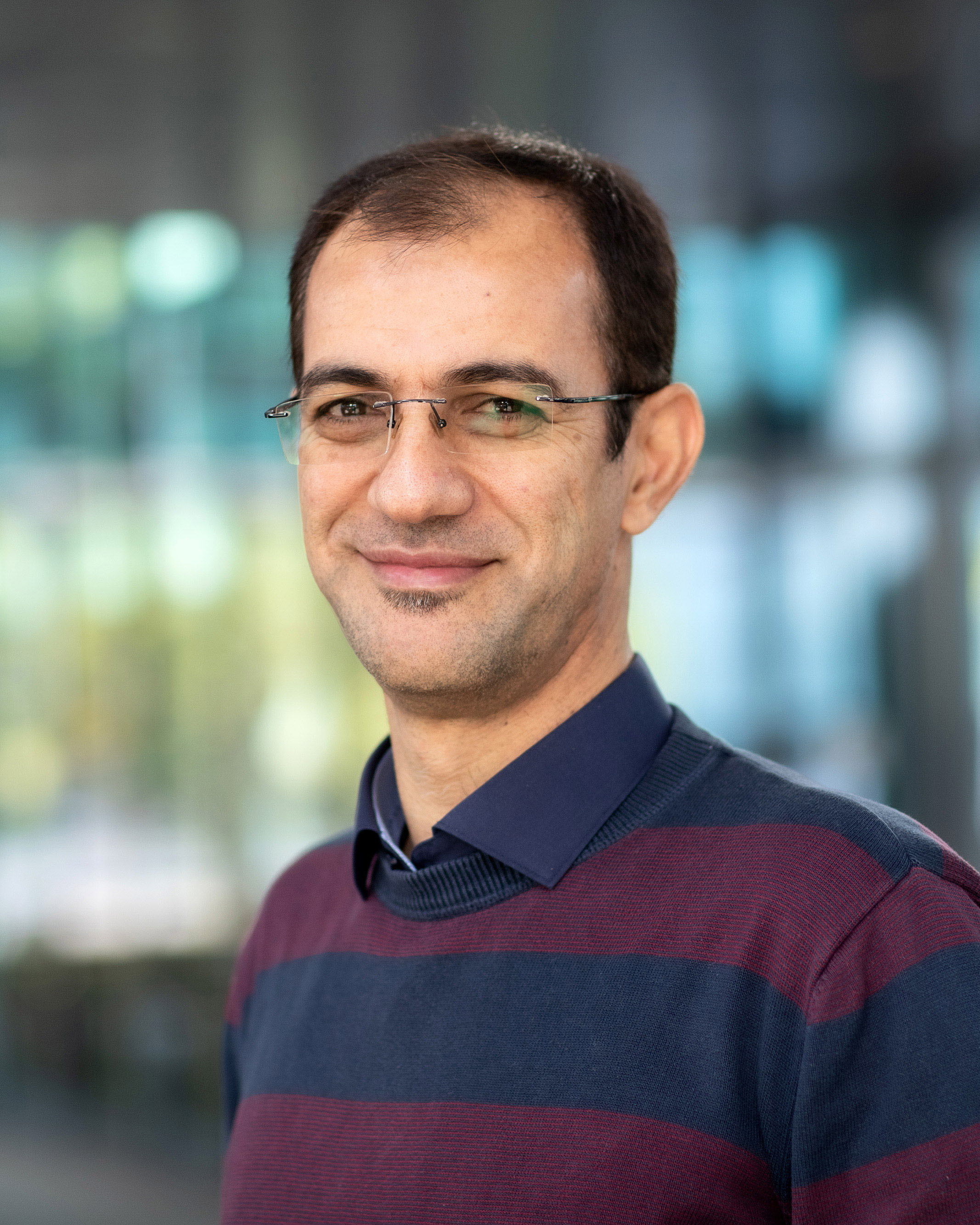}}]{Habib  Mostafaei} is currently an Assistant Professor of Computer Science at the Eindhoven University of Technology. He received the Ph.D. degree in Computer Science and Engineering from Roma Tre University in 2019. Before, he was a postdoctoral researcher at Technische Universität Berlin, where he was involved in the BIFOLD-BBDC project from 2019-2022. He worked as a full‐time faculty member at the Computer Engineering Department of Azad University from 2009‐2015. He is a member of ACM and IEEE. His main research fields include networked systems, network management, and distributed systems. For additional information: \url{https://mostafaei.bitbucket.io} 
\end{IEEEbiography}

\begin{IEEEbiography}[{\includegraphics[width=1in,height=1.25in,clip,keepaspectratio]{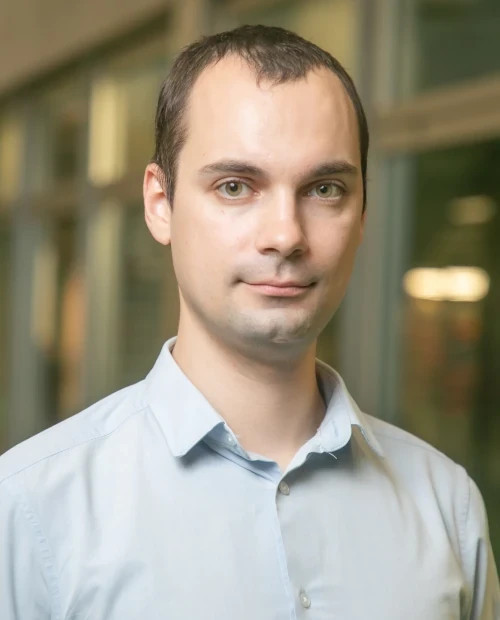}}]{Maciej Pacut} is a postdoctoral researcher at TU Berlin, Germany. He received his PhD from University of Wroclaw, Poland in 2019. Subsequently, Maciej Pacut worked as a postdoc at University of Vienna, Austria from 2020 to 2021. His research interests revolve around online algorithms for fundamental problems of computer networks and data centers.
\end{IEEEbiography}

\begin{IEEEbiography}[{\includegraphics[width=1in,height=1.25in,clip,keepaspectratio]{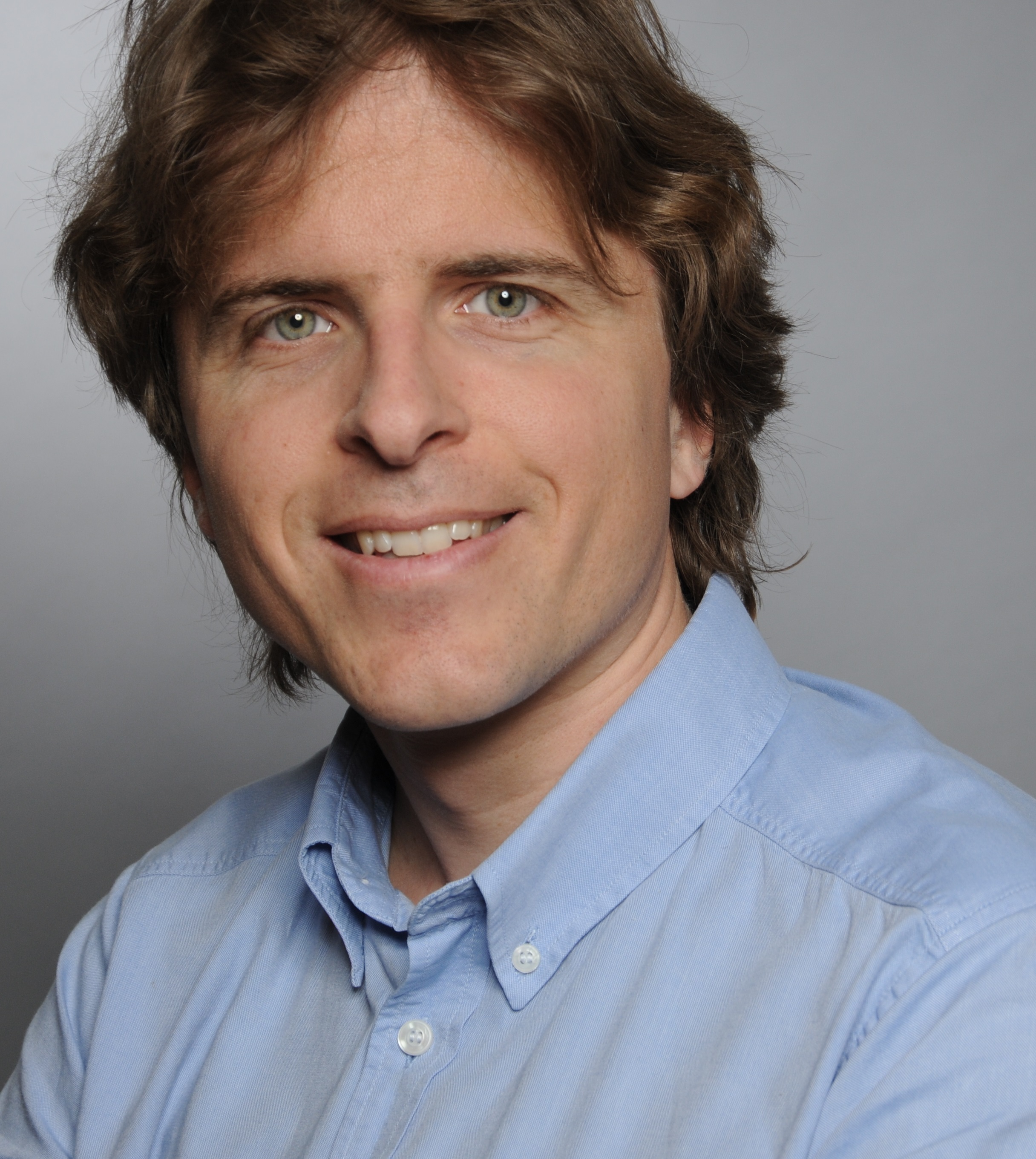}}]{Stefan Schmid} is a Professor at TU Berlin, Germany. He received his MSc (2004) and PhD (2008)
from ETH Zurich, Switzerland. Subsequently, Stefan
Schmid worked as postdoc at TU Munich and the
University of Paderborn (2009). From 2009 to 2015,
he was a senior research scientist at the Telekom Innovations Laboratories (T-Labs) in Berlin, Germany,
from 2015 to 2018 an Associate Professor at Aalborg University, Denmark, and from 2018 to 2021
a Professor at the University of Vienna, Austria.
His research interests revolve around algorithmic
problems of networked and distributed systems, currently with a focus on
self-adjusting networks (related to his ERC project AdjustNet) and resilient
networks (related to his WWTF project WhatIf).
\end{IEEEbiography}

\balance
\end{document}